# Unsupervised cryo-EM data clustering through adaptively constrained K-means algorithm

Yaofang Xu[1], Jiayi Wu[2], Chang-Cheng Yin[1]*, Youdong Mao[2,3]*

[1]Department of Biophysics, Peking University Health Science Center, Beijing 100191, China.

[2]State Key Laboratory of Artificial Microstructure and Mesoscopic Physics, Institute of Condensed Matter Physics, School of Physics, Center for Quantitative Biology, Peking University, Beijing 100871, China.

[3]Intel Parallel Computing Center for Structural Biology, Dana-Farber Cancer Institute, Department of Microbiology and Immunobiology, Harvard Medical School, Boston, MA 02115, USA.

*Corresponding authors

E-mail: youdong_mao@dfci.harvard.edu (YM)

E-mail: ccyin@hsc.pku.edu.cn (CY)

# Abstract

In single-particle cryo-electron microscopy (cryo-EM), K-means clustering algorithm is widely used in unsupervised 2D classification of projection images of biological macromolecules. 3D *ab initio* reconstruction requires accurate unsupervised classification in order to separate molecular projections of distinct orientations. Due to background noise in single-particle images and uncertainty of molecular orientations, traditional K-means clustering algorithm may classify images into wrong classes and produce classes with a large variation in membership. Overcoming these limitations requires further development on clustering algorithms for cryo-EM data analysis. We propose a novel unsupervised data clustering method building upon the traditional K-means algorithm. By introducing an adaptive constraint term in the objective function, our algorithm not only avoids a large variation in class sizes but also produces more accurate data clustering. Applications of this approach to both simulated and experimental cryo-EM data demonstrate that our algorithm is a significantly improved alterative to the traditional K-means algorithm in single-particle cryo-EM analysis.

# Introduction

Single-particle reconstruction in cryo-electron microscopy (cryo-EM) is a powerful technology to determine three-dimensional structures of biological macromolecular complexes in their native states (1). Recent advances in electron-counting detector and high-performance computing enabled 3D structural determination of biological macromolecular complexes at a near-atomic resolution (2-4). The goal of single-particle reconstruction is to recover the 3D structure of a macromolecule from a large number of 2D transmission images, in which the macromolecule assumes random, unknown orientations.

Due to high sensitivity of biological samples to radiation damage by electron beam, cryo-EM data are often acquired with very limited electron doses (10-50 electron/Å$^2$), which makes the cryo-EM images extremely noisy. To determine the relative orientations of molecular projections, a crucial step is to classify 2D projection images in an unsupervised fashion such that images in the same class come from similar orientations of projection (5, 6). For each class, images are aligned, centered and averaged to produce class averages with enhanced signal-to-noise ratio (SNR). Generating 2D class averages is important for both common-line based 3D *ab initio* reconstruction (7-14) and some modern methods (15, 16). Unsupervised classification is also useful for a quick evaluation on structural heterogeneity and quality of samples before entering time-consuming 3D refinement steps (17, 18).

If ignoring conformational dynamics of imaged macromolecules, the intrinsic difference among projection images mainly comes from two sources: projection direction and in-plane rotation. Prior to classification, single-particle images must be aligned to minimize the differences in their translation and in-plane rotation. There are two popular approaches for initial classification of 2D projection images, namely, multi-reference alignment (MRA) (19) and reference-free alignment (RFA) (20). In MRA, a 2D image alignment step and a data-clustering step are performed iteratively until convergence. In the 2D image alignment step, each image is rotated and shifted incrementally with respect to each reference. All possible correlations between the rotated, translated image and the reference are computed. The distance between an image and a reference is defined as the minimum of all correlation values between them. Based on these distances, in the data-clustering step, traditional K-means clustering is used to classify all images into different classes. An implementation of the MRA strategy is found in SPARX (21). In RFA, all images are first aligned globally, which attempts to find rotations and translations for all images that minimize the sum of squared deviation from their mean. These aligned images are used as the input for data-clustering algorithms. This strategy was implemented in SPIDER (22).

Moreover, upon the suggestion of Jean-Pierre Bretaudière, multivariate statistical analysis (MSA) was introduced into cryo-EM (6, 23, 24). MSA reduces the dimensionality of images by projecting them into a subspace spanned by several eigenvectors, which are also called features. Reducing dimensionality not only accelerates computing but also denoises projection images. The resulting features can also be used as the references for image alignment. For example, EMAN2 combines MSA with MRA (MSA/MRA) (see its script e2refine2d.py) (25). It first generates translational and rotational invariants for initial classification. Then, a MSA step is iterated with a MRA step, in which images are aligned to those features and classified by the K-means algorithm, until a pre-defined number of iterations is reached.

Despite rapid progress in data science, the traditional K-means algorithm remains one of the most popular data clustering approaches for single-particle cryo-EM. However, the traditional K-means algorithm has certain limitations. A class average with a higher SNR correlates preferably with noise in the high-frequency domain, resulting in attraction of more images into high-SNR classes. Therefore, when used with MRA, traditional K-means clustering tends to misclassify single-particle images to classes with more members (26). Moreover, some classes may be depleted during iterations. Reseeding empty classes may tentatively remedy this problem. However, it can also break the balance of class sizes among distinct classes, resulting in the coexistence of both oversized and undersized classes. The same issue was also found in multi-reference maximum-likelihood classification, implemented in XMIPP and RELION (27-29).

To avoid class size getting improperly large, several approaches were proposed (22, 26, 27, 30). First, a modified traditional K-means was implemented in SPIDER for data clustering, where the objective function is multiplied with a factor $\sqrt{s/(s \pm 1)}$. Here, s is class size. '-' is adopted when an image is compared with its own class; and '+' is adopted when the image is compared to other classes. For very large classes, this factor is almost 1, whereas it is well below one for small classes. Therefore, it tends to classify more images to small classes. This modification does avoid empty class in the traditional K-means, but it does not exclude over-sized classes or ones containing only one image (We refer this modified traditional K-means in SPIDER as the traditional K-means in later context, when there is no ambiguity.). Second, an algorithm called equal-sized group K-means (EQK-means) was developed to control the sizes of classes (30), which was implemented in SPARX. In each iteration of EQK-means, every class is forced to have the same number of image members, which avoids the attraction of images to high-SNR classes. Therefore, the resulting class averages may achieve comparable SNRs. However, given that the experimental projection directions cannot be absolutely evenly distributed, it is problematic to produce equally sized classes. Classes in denser angular areas should be assigned with more image members than classes in sparser areas. Third, a modified MRA approach using the CL2D algorithm (26) was implemented in XMIPP (29), which classifies images hierarchically. At each hierarchical level, images are classified with a control of class size by dividing large classes into more classes. The hierarchical approach conducts classification at each level and may require more CPU time than non-hierarchical approaches.

In this study, we introduce a novel data clustering approach, named adaptively constrained K-means algorithm (ACK-means), for unsupervised cryo-EM image classification. Different from EQK-means that enforces an equal size on all classes, ACK-means controls the class size with an adaptive balance between class size and classification accuracy. Thus, ACK-means can in principle avoid an excessive growth of class size while producing a more accurate angular assignment. Our study suggests that ACK-means is a significantly improved alternative to the traditional K-means for cryo-EM data clustering.

# Methods

## A brief review on the traditional K-means algorithm

Let $X = \{x_1, x_2, \ldots, x_n\}$ represent a set of projection images to be classified. Each class is represented by a centroid $\mu_j$, for $j = 1, 2, \ldots, k$. The goal is to partition $X$ into $k$ classes so that the following objective function is minimized:

$$J_k = \sum_{j=1}^{k} \sum_{p_i=j} dissim(x_i, \mu_j),$$

where $dissim(\cdot,\cdot)$ is a function measuring the dissimilarity between images $x_i$ and $\mu_j$. The partitioning result is denoted by an assignment vector $\boldsymbol{p} = (p_1, p_2, \ldots, p_n)$, which assigns image $\boldsymbol{x}_i$ to the $p_i$-th class. The most commonly used dissimilarity measure is the Euclidian distance (31). In most MRA approaches, the dissimilarity is defined as a minimum over all possible relative rotations and translations of an image with respect to another.

To solve this minimization problem globally is NP-hard (32). As a local minimum solution, the traditional K-means algorithm was first developed by MacQueen (33). He gave the name "K-means" to the algorithm that assigns each image to the class of the nearest centroid. This can be formulated in the following:

(1) **Initialization step**: Determine $k$ initial centroids (seed points) $\{\boldsymbol{\mu}_j\}_{j=1}^k$ by randomly selecting $k$ images from $\boldsymbol{X}$;

(2) **Assignment step**: For each image, assign it to the class specified by the most similar centroid;

$$p_i = \arg\min_j dissim(\boldsymbol{x}_i, \boldsymbol{\mu}_j)$$

(3) **Update step**: Recalculate the centroid by the image mean in each class:

$$\boldsymbol{\mu}_j = mean(\{\boldsymbol{x}_i : p_i = j\})$$

(4) Repeat (2) and (3) until there are no more changes of membership.

## Adaptive constraint

To introduce an adaptive constraint to K-means clustering, we add an additional term to the objective function as shown in the following expression:

$$J = \sum_{j=1}^{k} \sum_{p_i=j} dissim(x_i, \mu_j) + \lambda ss^{T}. \qquad (1)$$

The first term is the sum of dissimilarity between image $x_i$ and centroid $\mu_j$. In the second term, $s$ is a vector, whose element $s_i$ denotes the number of images belonging to the $i$-th class; and $\lambda$ is a non-negative parameter.

Note that the sum of all the elements of $s$ is the total number of images, which is a constant n. According to Cauchy-Schwarz inequality, the second term is minimized only when $s_1 = s_2 = \cdots = s_k$. Therefore, by introducing the second term, we establish a competition between dissimilarity and the balance of class sizes. If we set $\lambda = 0$, no constraint is exercised on class size and minimizing (1) is exactly the same as the traditional K-means algorithm. If $\lambda = +\infty$, all classes would have the same size. As $\lambda$ is changed from 0 to $+\infty$, more weight is given on the balance of class size. It allows us to tune the class sizes adaptively by regulating the strength of the constraint, as opposed to the EQK-means algorithm that enforces equally sized classes (30). For this reason, we call the second term an adaptive constraint. Thus, the proposed algorithm allows more images to update their membership in a large class than in a small class during the optimization of the objective function.

For a given set of centroids, $n$ images are assigned to $k$ classes one by one through minimizing the objective function (1). Suppose at the end of the previous iteration, image $x_i$ is assigned to class $p_i$ and the class size vector is $s = (s_1, s_2, \ldots, s_k)$. Then, in the current iteration, the class size vector is first recalculated as $s'$ by omitting $x_i$. Almost all the elements of $s'$ is the same as $s$, except $s'_{p_i} = s_{p_i} - 1$. Then $x_i$ is reassigned to class $p_i$ by solving the following minimization:

$$p_i = \arg\min_j \{ dissim(x_i, \mu_j) + \lambda \sum_{h=1}^{k} [s'_h + \delta_{hj}]^2 \}$$

$$= \arg\min_j \{ dissim(x_i, \mu_j) + 2\lambda s'_j + \lambda + \lambda \sum_{h=1}^{k} s'^2_h \}$$

$$= \arg\min_j \{ dissim(x_i, \mu_j) + 2\lambda s'_j \},$$

where $\delta_{hj}$ is the Kronecker delta function. When all the images are reassigned, we end up with a new assignment vector $p$ describing the partition in the current iteration. For the given partitioning, to further minimize the objective function (1), we update the $k$ centroids by averaging images in the same class. This process is iterated until there are no more changes in membership.

## Characteristic dissimilarity

Due to background noise and variation in molecular projections, the scale of pixel intensities in single-particle images is expected to vary from case to case. To keep the competition between the two terms of equation (1) at the same magnitude, one needs a larger $\lambda$ for an image dataset with large dissimilarities than that with small dissimilarities. Therefore, we developed a strategy to tune the value of $\lambda$ that is applicable to varying scales. One quantity reflecting data scaling is the maximum value of dissimilarity between any pairs of images in a given dataset. However, computing all the dissimilarities between any image pair is extremely time-consuming and practically prohibited. Instead, because we already computed dissimilarities between images and centroids during image assignment to different classes, we can construct a quantity called "characteristic dissimilarity" from these values. In each iteration, we randomly select 10 images. For each one of them, we find its smallest and largest dissimilarities among the k dissimilarities with k centroids. The difference between the largest and smallest dissimilarities is calculated for each image, and is then averaged together to make characteristic dissimilarity $d_c$. Hence, we rewrite $2\lambda$ as:

$$2\lambda = \beta \frac{d_c}{\lfloor n/k \rfloor},$$

where $\beta$ is a free parameter in the range from 0 to $+\infty$. So we decompose $2\lambda$ into two parameters. The first parameter $d_c$ describes the change of pixel intensity scaling, whereas the second parameter $\beta$ decides the weights on class size whose value is independent of data scaling. The constant $\lfloor n/k \rfloor$ is the class size if all images are partitioned equally. Hence, if the partition of all classes are ideally balanced, $\Delta s'_j / \lfloor n/k \rfloor$ represents the fraction of changed images in the $j$-th class. Given $0 < \beta < +\infty$, images are partitioned according to dissimilarity while class size is monitored by the adaptive constraint. Note that the only parameter to be considered during the application of ACK-means is $\beta$. The smaller $\beta$ is, the less we consider the balance of class sizes. Our experiments show that ACK-means can generate satisfying results with $\beta = 0.5$ (see below). In this case, if $d_c$ is the diameter of the area occupied by the data, then $\beta d_c$ is the radius.

## Implementation algorithm

The algorithm of adaptively constrained K-means is implemented in the following pseudo code.

| **Algorithm: ACK-means** |
|---|
| **Input:** |
| $\{x_i\}_{i=1}^{n}$: Set of data points. |

$k$ : Number of classes.

$\sigma_0$: The minimum fraction of data points that are changed membership.

$\beta$: The weight on the adaptive constraint term.

1: Initialize centroids $\{\mu_j\}_{i=1}^{k}$ by randomly selecting $k$ data points from input data.

2: Compute $p$ according to $p_i = \arg\min_{j} dissim(x_i, \mu_j)$.

3: Update $\{\mu_j\}_{i=1}^{k}$ by averaging data points in the same class.

4: **while** $\sigma > \sigma_0$ **do**

5:     Randomly select 10 images $\{r_m\}_{m=1}^{10}$ and compute $t_m$ for $m=1:10$ according to:

$$t_m = \max_{j} dissim(r_m, \mu_j) - \min_{j} dissim(x_m, \mu_j)$$

6:     Compute $d_c$ according to $d_c = \sum_{m=1}^{10} t_m / 10$

7:     Compute $\lambda$ according to $\lambda = \beta \frac{d_c/2}{[n/k]}$.

8:     Save the old assignment vector according to $p_{old} = p$

9:     **for** $i = 1 : n$ **do**

10:        Compute $s'$ according to $s'_l = \sum_{m \neq i} \delta_{p_m, l}$

11:        Update $p_i$ according to $p_i = \arg\min_{j} dissim(x_i, \mu_j) + 2\lambda s'_j$

12:     **end for**

13:     Update $\{\pmb{\mu}_j\}_{i=1}^{k}$ by averaging data points in the same class.

14:     Compute $\pmb{\sigma}$ according to $\pmb{\sigma} = \mathbf{1} - \frac{1}{n}\sum_{i=1}^{n}\pmb{\delta}_{p_i^{old},p_i}$

15: **end while**

Return: assignment vector $\pmb{p}$.

---

Note that we exercise no assumption regarding classification and initialize the algorithm as what the traditional K-means does. To ensure an unsupervised nature of the classification, $k$ images are selected randomly from the dataset as the initial centroids. This guarantees that there is at least one member in each class. Then, each of the rest images is assigned to the class whose centroid is the nearest to the image. To devise a termination criterion for the algorithm, we set a threshold parameter $\sigma_0$ here. If the number of data points changing their membership in the current iteration decreases to this threshold, the algorithm is terminated.

## Benchmark with simulated data

The density map of Escherichia coli 70S ribosome (34) was used to generate 10,000 simulated projection images (S1 Fig). In reality, most protein structures are of lower symmetry or asymmetric. Therefore, some orientations are expected to appear more frequently than others in vitreous ice. In order to emulate this phenomenon, we uniformly chose 100 orientations covering half a sphere. Each orientation is regarded as a Gaussian center, around which 100 projections were generated with a Gaussian distribution. Due to electron lens aberrations and defocusing, we further modified the projection images with the contrast transfer function (CTF). The projections were then additively contaminated with Gaussian noise at different SNR = 1/3, 1/10, 1/30 (S2 Fig), which allowed us to investigate the proposed algorithm at different noise level. The input projections to all experiments were CTF-corrected by phase flipping (35).

To examine the performance of our algorithm, we compared the results of classifying the 10,000 simulated images into 100 classes by using ACK-means with those from other existing approaches. For the standard MRA, we compared ACK-means against traditional K-means and EQK-means algorithms implemented in SPARX. The script isac.py in SPARX is part of a method called ISAC (Iterative Stable Alignment and Clustering) proposed in (36), consisting of the standard MRA part, followed by analysis within classes. The within-class analysis traces the change of membership of images and selects stably classified images that do not change their membership in each iteration. To focus on the effect of different data clustering algorithms, only the MRA part is used in our test. To see the influence of ACK-means in MRA with MSA (MRA/MSA), we replaced the traditional K-means with ACK-means of e2refine.py in EMAN2 and compared their performance. For RFA, we followed the protocol of SPIDER (34) and compared the classification results of the traditional K-means (precisely, it is a modified traditional K-means, but we refer it as the traditional K-means without ambiguity) with those of ACK-means.

# Results

## Simulated data

Since all the original angles of the input projections are known, the angular difference between any pair of projections in each class, also termed "angular distances", can be computed. The statistical behavior of the angular distances can be used to measure the quality of the corresponding class (26). A class assigned with *n* image members has $\binom{n}{2}$ pairs of angular distances. We used two plots to compare the results from different algorithms. The first plot is the histogram distribution of angular distances from all the classes (26). A better algorithm is expected to exhibit a distribution curve with a sharper, higher peak at lower angular distances. The second plot ranks the sizes of all classes to show the balance of classification. As shown in Fig 1, we compared the unsupervised classification results by ACK-means on the simulated with a SNR of 0.1 with those obtained by several existing K-means implementations: the traditional K-means and EQK-means in the standard MRA approach implemented in SPARX (21, 30), the traditional K-means in the MRA/MSA approach implemented in EMAN2 (25), and the one in the RFA approach implemented in SPIDER (22).

In the MRA approach, although EQK-means avoids the attraction of dissimilar images by delivering equally sized classes, it exhibits reduced angular accuracy of classification (Fig 1a, Fig 1b). By contrast, ACK-means makes little compromise on the balance of class size, yet improves the classification accuracy (Fig 1a, Fig 1b). In both MRA and MRA/MSA approaches, ACK-means gives rise to a prominent improvement in both the classification accuracy and the balance of class size (Fig 1a-d). However, in the RFA approach, although the improvement of classification accuracy is not obvious (Fig 1e), ACK-means still generate a more balanced class size (Fig 1f).

The three experiments behave differently as SNR is changed (S3 and S4 Figs). Since MRA has the strongest effect of attraction of dissimilar particles over other approaches, the attraction effect becomes much stronger with decreasing SNR. By contrast, the performance of ACK-means in controlling the class size adaptively does not degrade with decreasing SNR (Figs 1a and 1b; S3a, S3b,

S4a and S4b Figs). In the MRA/MSA approach, ACK-means outperforms the traditional K-means at moderately low SNR level (Fig 1c and S3c Fig). However, at lower SNR (0.033), their difference in the histogram disappeared (S4c Fig). This is likely because alignment errors introduced in the early step cannot be eliminated by ACK-means in the later step. In the RFA approach, ACK-means generates similar classification accuracy with the traditional K-means at all noise levels (Fig 1e, S3e and S4e Figs). This result confirms that the classification accuracy is bound by the alignment error. In all cases, ACK-means gave rise to a well-balanced classification (S4d Fig).

## Experimental cryo-EM data

Three real experimental datasets were used to examine our ACK-means algorithm in this study. We compared the results of our algorithm against the traditional K-means implementations in SPARX, EMAN2 and SPIDER, as well as EQK-means in SPARX. Since it is impossible to know the true projection angles of individual single particles, we evaluate the classification results by inspecting the quality of 2D class averages visually.

### Case 1: GroEL

The first dataset consists of 5,000 particles selected from a GroEL dataset of 26 micrographs, whose pixel size is 2.10 Å/pixel. The size of particles is 140 × 140 pixels (37). These particles were first phase-flipped and then classified into 25 classes by different algorithms. As shown in Fig 2, a set of 2D class averages were computed with the traditional K-means (Fig 2a), EQK-means (Fig 2b) and ACK-means (Fig 2c) through the MRA protocol in SPARX. The traditional K-means cannot monitor class sizes and produced more blurred classes than other two algorithms (Fig 2a). Although EQK-means and ACK-means both produced balanced results, EQK-means generated the two worst class averages among all class averages (Fig 2b). For the MRA/MSA approach in EMAN2, class averages from the traditional K-means (Fig 3a) and ACK-means (Fig 3b) were compared. ACK-means generated clearer class averages with balanced sizes. Furthermore, comparison between the traditional K-means (Fig 4a) and ACK-means (Fig 4b) were made with the RFA approach in SPIDER. Both the traditional K-means and ACK-means generated class averages of comparable quality, but the later substantially improved the balance of class sizes, avoiding both oversized and empty classes.

**Case 2: Inflammasome**

We used 281 particles of inflammasome to benchmark our algorithm (38). The data were collected with a pixel size of 1.72 Å/pixel and an acceleration voltage of 200 kV. The particles have a size of 160 × 160 pixels. These particles were pre-selected such that only side views with different lengths, corresponding to different oligomeric states of inflammasome, were included in the dataset. After phase-flipped, the dataset was classified into 20 classes using different algorithms. In the MRA approach with SPARX, the class averages were generated by the traditional K-means (Fig 5a), EQK-means (Fig 5b) and ACK-means (Fig 5c). Without the constraint on class sizes, many classes in the traditional K-means were not assigned with enough particles, producing blurred class averages (Fig. 5a). Although many class averages of EQK-means and ACK-means are similar, some classes of EQK-means present misaligned features, indicating the failure of classifying different particles. Similarly, when compared in the MRA/MSA approach with EMAN2, ACK-means produced generally improved classification results than the traditional K-means (Figs 6a and 6b). We further compared our approach with the traditional K-means in the RFA approach implemented in SPIDER. The traditional K-means generated many classes with only one particle (Fig 7a). By contrast, this was well avoided in the results from unsupervised classification by our ACK-means algorithm (Fig 7b).

## Case 3: Proteasome

Finally, we used a dataset containing proteasomal RP (regulatory particle) and RP-CP (regulatory particle associated with core particle) subcomplex (39). The total number of particle is 3,960, with pixel size 2.00 Å/pixel and particle size 160 × 160 pixels. All particles are pre-processed by phase-flipping and classified into 40 classes. For the MRA approach in SPARX, the class averages generated by the traditional K-means, EQK-means and ACK-means are shown in Figs 8a-8c, respectively. The traditional K-means (Fig 8a) generated many blurred classes as a result of no constraint on class sizes. ACK-means (Fig 8c) and EQK-means (Fig 8b) have comparable results. For the MRA/MSA approach in EMAN2 and RFA approach in SPIDER, we compared the class averages of traditional K-means (Figs 9a and 10a) and ACK-means (Figs 9b and 10b). Class averages of ACK-means show more classes with clear details. Additionally, the traditional K-means in SPIDER generated 3 classes with only one particle.

## DISCUSSION

In this study, we propose a new data-clustering algorithm, which generates adaptively balanced, unsupervised classification, preventing the attraction of dissimilar particles into classes of large sizes or higher SNRs. This allows significant improvement in unsupervised image classification over the traditional K-means algorithm. Meanwhile, by controlling class sizes adaptively, our approach also improves angular accuracy of image clustering as compared to EQK-means, allowing more particles to be assigned to a class if the operation can improve classification accuracy.

We tested our algorithm with both simulated and real experimental datasets. The projection orientations of simulated data were generated with both dense and sparse angular areas to imitate realistic situation. We found that our ACK-means algorithm consistently outperforms the traditional K-means in all cases. In MRA, traditional K-means suffers from attracting dissimilar particles into classes with more particles. It does not control class sizes and often generates many classes of very few image members, resulting in blurred class averages. ACK-means and EQK-means both generate balanced class averages, whereas ACK-means gives rise to improved classification accuracy, allowing more details recovered in class averages. In contrast to that EQK-means avoids the growing of class sizes by forcing each class to have the same size, ACK-means monitors class sizes adaptively when determining the class assignment of particles.

In the tests with the simulated and experimental GroEL data, we found little improvement on the accuracy of classification by ACK-means against the traditional K-means in the RFA approach in SPIDER. However, in the test with the experimental inflammasome and RP data, we observed prominent differences between the two algorithms. Although the traditional K-means was modified by a factor in SPIDER, it still generated many classes with only one image. By contrast, ACK-means produced balanced classes with more informational class averages. Interestingly, we further combined ACK-means with our recently proposed statistical manifold learning algorithm (39) and found a significant improvement in the RFA approach (data not shown). It bodes well for the future development of improved data clustering protocols that integrate both ACK-means and manifold learning approaches. In summary, the ACK-means takes into account both the classification accuracy and the balance of class sizes. It presents a significantly improved alternative to the traditional K-means as a data-clustering algorithm for cryo-EM analysis.

# Acknowledgements

The authors thank D. Yu, Y. Zhu, Y. Wang, and Q. Ouyang for helpful discussion.

# Funding

This work was funded by a grant of the Thousand Talents Plan of China (Y.M.), by a grant from National Natural Science Foundation of China 91530321 (Y.M.), by the Intel Parallel Computing Center program (Y.M.). The cryo-EM experiments were performed in part at the Center for Nanoscale Systems at Harvard University, a member of the National Nanotechnology Coordinated Infrastructure Network (NNCI), which is supported by the National Science Foundation under NSF award no. 1541959. The cryo-EM facility was funded through the NIH grant AI100645, Center for HIV/AIDS Vaccine Immunology and Immunogen Design (CHAVI-ID). The data processing was performed in part in the Sullivan supercomputer, which is funded in part by a gift from Mr. and Mrs. Daniel J. Sullivan, Jr.

*Conflict of Interest:* none declared.

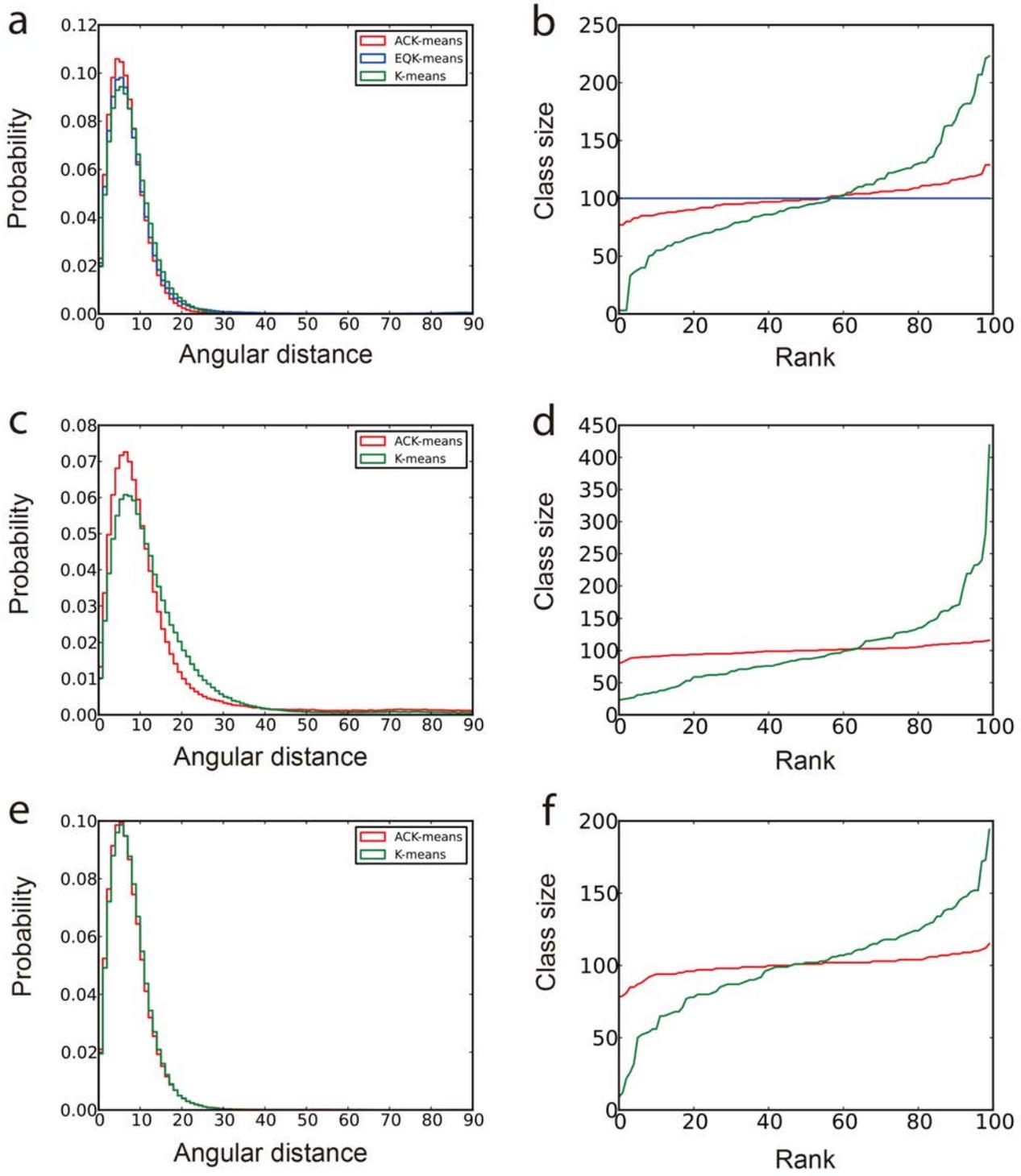

**Fig 1. Comparison of classification results of simulated data with SNR = 1/10.**

The First column (panels a, c and e) is the normalized histogram of angular distances. More accurate classification produces curve with higher peak concentrated at lower angular distance. The second column (panels b, d and f) shows the class sizes arranged in an ascend order. The most balanced classification has a horizontal line in this plot. (a) and (b) are for experiments using different clustering algorithms in MRA approach under SPARX. (c) and (d) are for experiments using different clustering algorithms in MRA/MSA approach under EMAN2. (e) and (f) are for experiments using different clustering algorithms in RFA approach under SPIDER. In all graphs, red curves present the results from ACK-means algorithm.

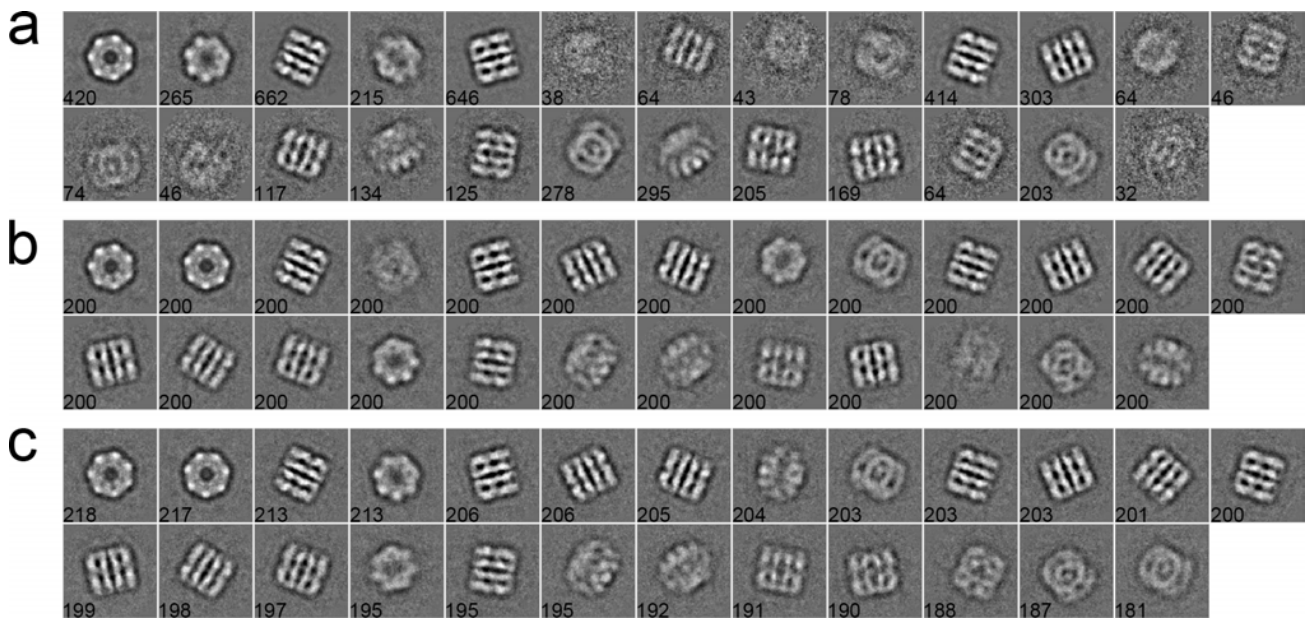

**Fig 2**. **2D class averages of GroEL using the traditional K-means (a), EQK-means (b) and ACK-means (c) in MRA approach from SPARX**.

Class size is shown at the left bottom of each class average. ACK-means (b) is the best by having the most number of clear classes.

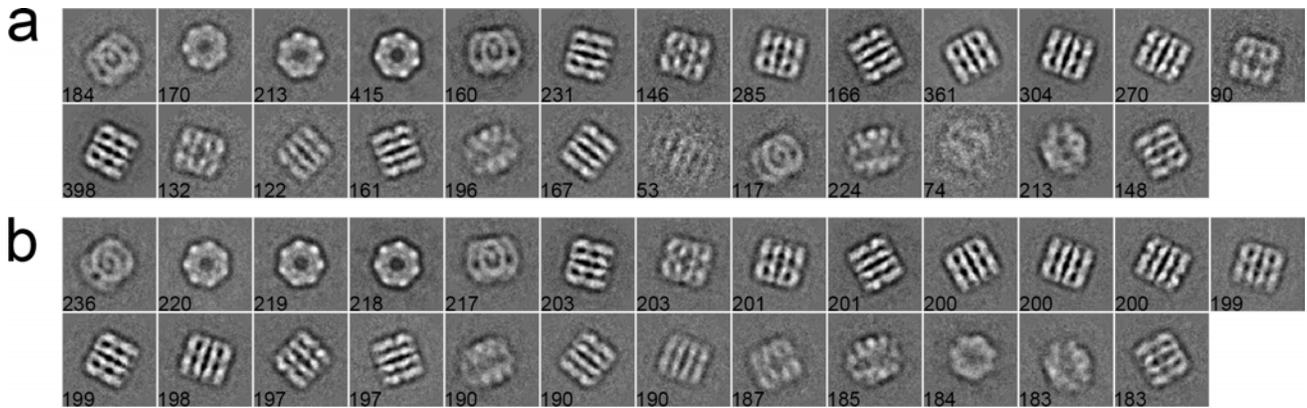

**Fig 3. 2D class averages of GroEL using the traditional K-means (a) and ACK-means (b) in MRA/MSA from EMAN2.**

Class size is shown at the left bottom of each class average. Their performance is similar, but ACK-means (b) has the more number of clear classes.

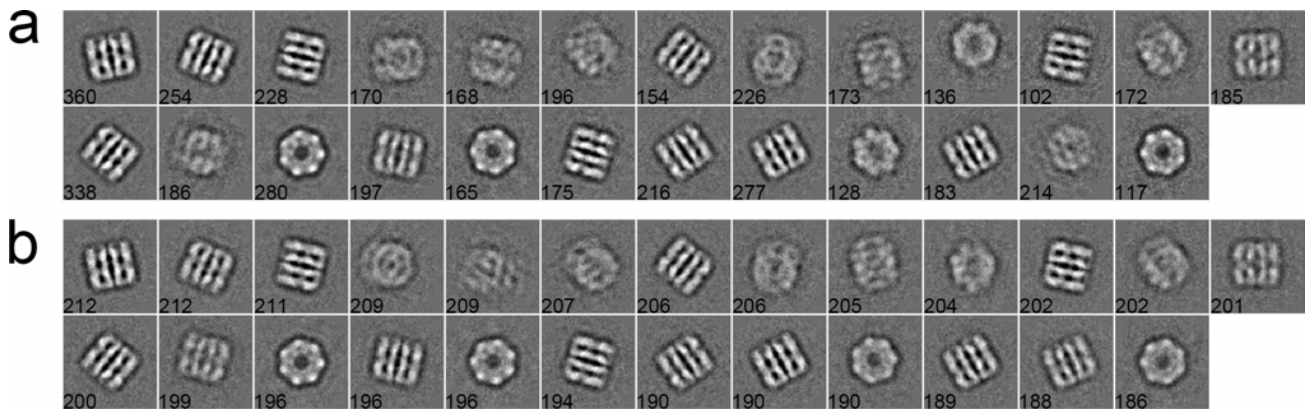

**Fig 4**. **2D class averages of GroEL using the traditional K-means (a) and ACK-means (b) in RFA from SPIDER.**

Class size is shown at the left bottom of each class average. The quality of class averages from both algorithms is comparable, but ACK-means (b) substantially improved the balance of class sizes.

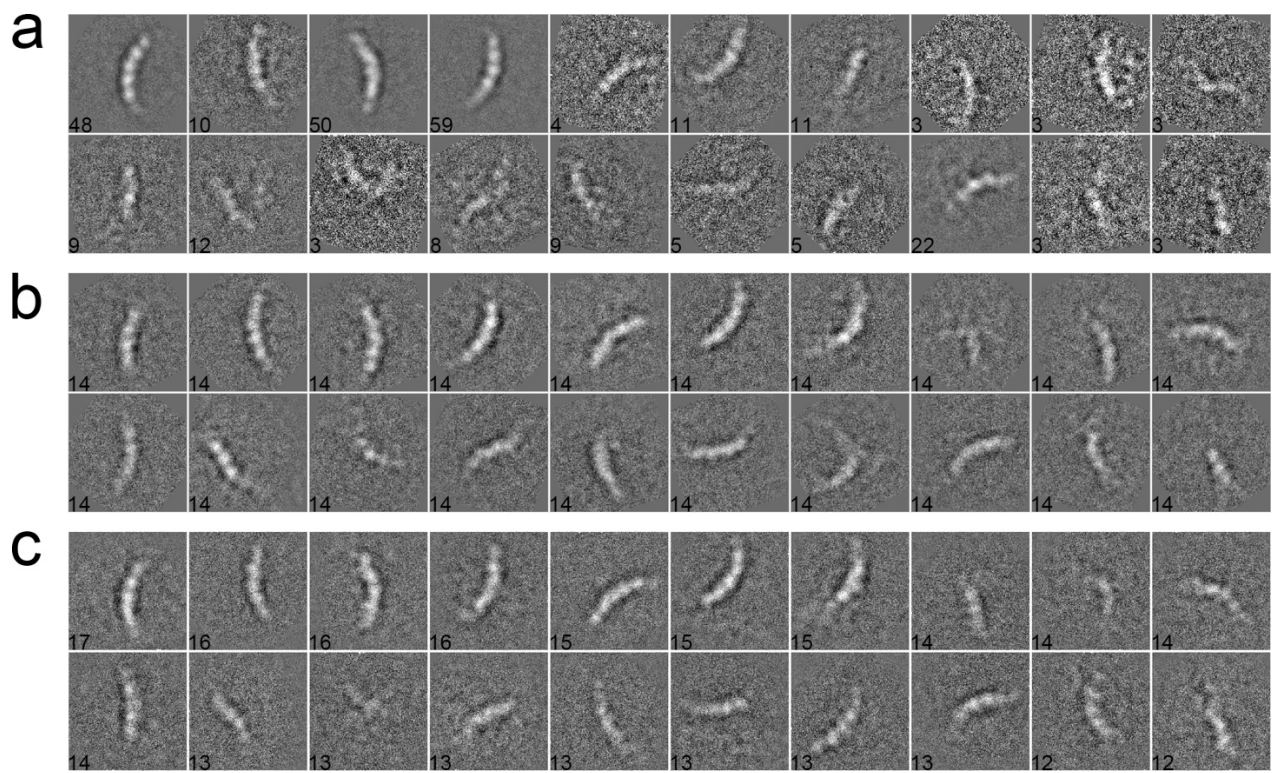

**Fig 5**. **2D class averages of Inflammasome using the traditional K-means (a), EQK-means (b) and ACK-means (c) in MRA from SPARX.**

Class size is shown at the left bottom of each class average. The traditional K-means generated many blurred class averages and EQK-means produced some class averages with misaligned features.

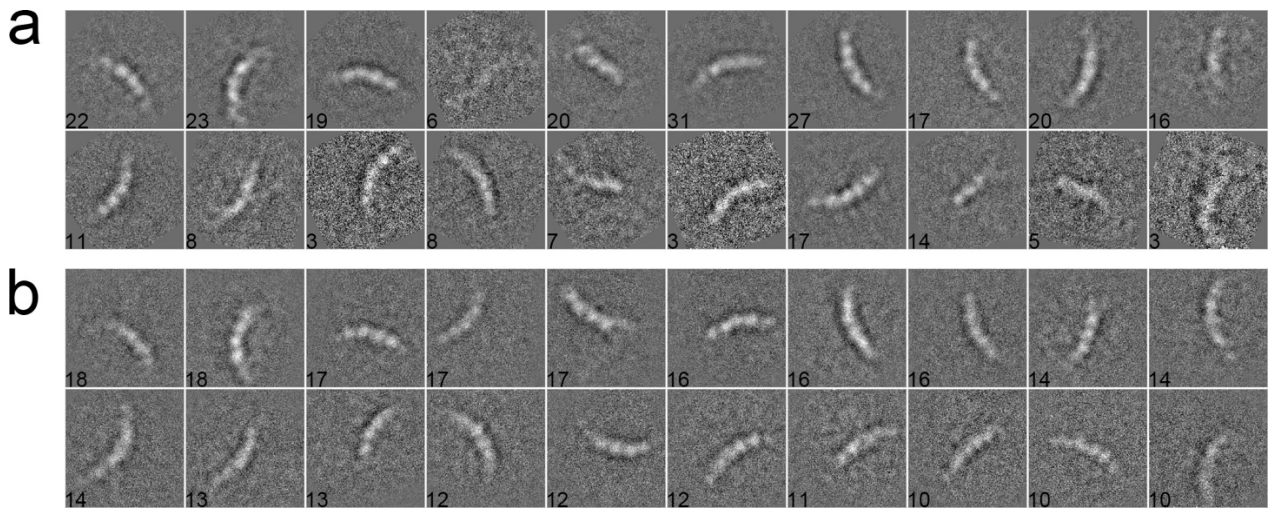

**Fig 6**. **2D class averages of Inflammasome using the traditional K-means (a), and ACK-means(b) MRA/MSA from EMAN2.**

Class size is shown at the left bottom of each class average. ACK-means generated improved results as compared to the traditional K-means.

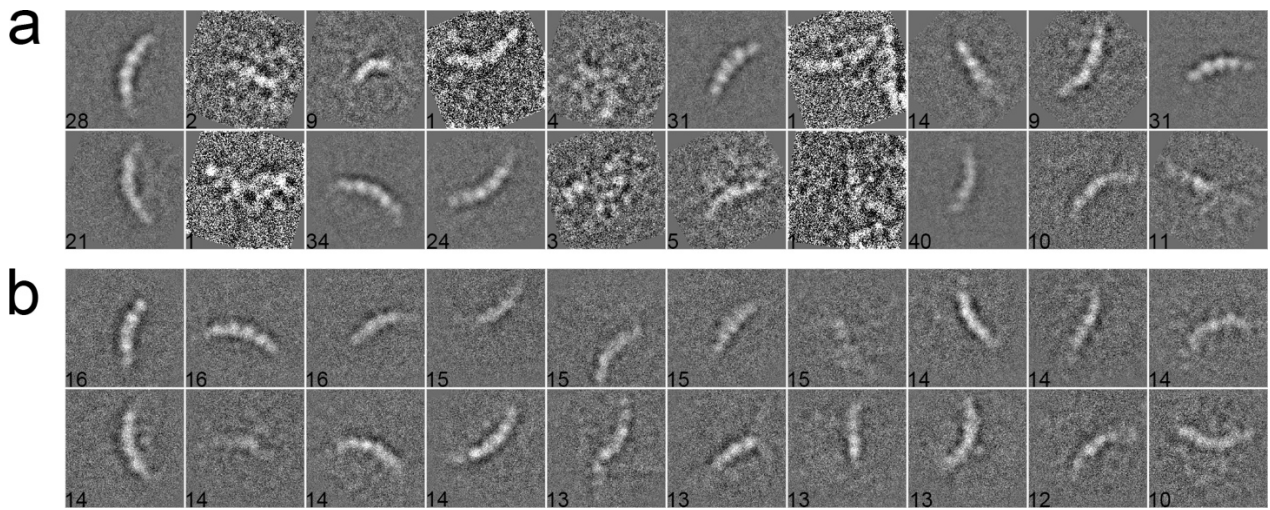

**Fig 7**. **2D class averages of Inflammasome using the traditional K-means (a), and ACK-means(b) in RFA from SPIDER.**

Class size is shown at the left bottom of each class average. There are many classes in (a) with only one particle. Traditional K-means generated many classes with only one particle, which is avoided in the results from ACK-means.

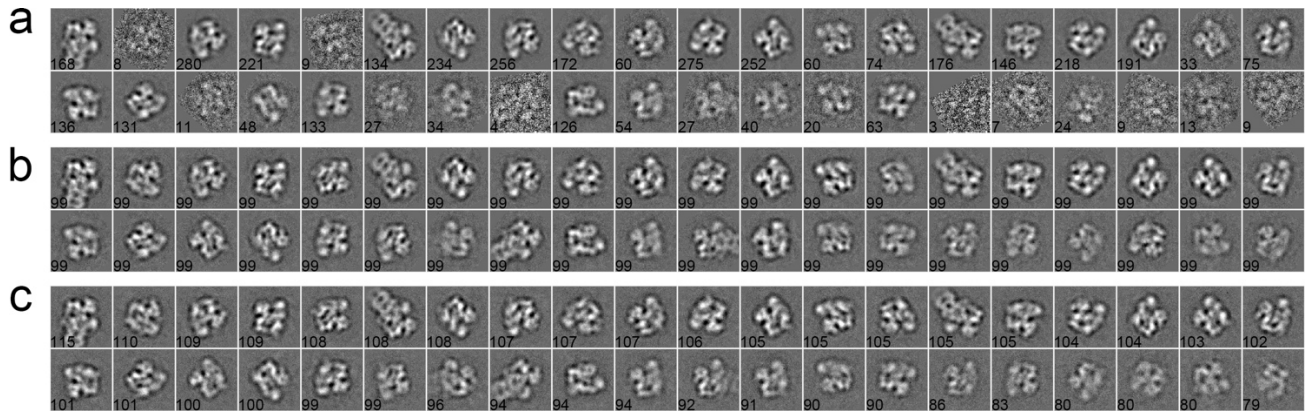

**Fig 8**. **2D class averages of RP using the traditional K-means (a) and ACK-means (b) in MRA from SPARX.**

Class size is shown at the left bottom of each class average. There are many blurred classes in (a) generated by the traditional K-means.

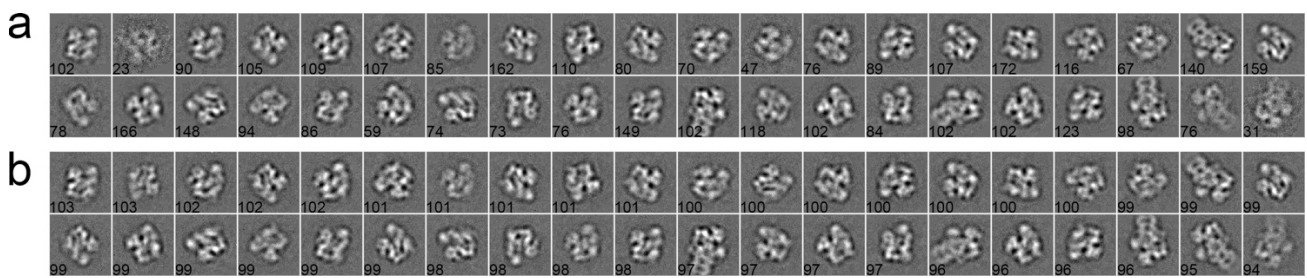

**Fig 9**. **2D class averages of RP using the traditional K-means (a) and ACK-means(b) in MRA/MSA from EMAN2.**

Class size is shown at the left bottom of each class average. Classes generated by ACK-means (a) are clearer than by the traditional K-means (a).

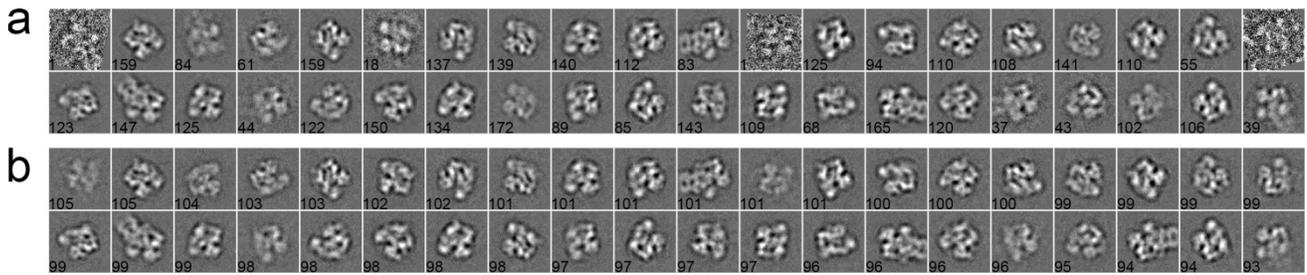

**Fig 10**. **2D class averages of RP using the traditional K-means (a) and ACK-means(b) in RFA from SPIDER.**

Class size is shown at the left bottom of each class average. The traditional K-means (a) generated some poor classes with only one particle. The performance of ACK-means (b) is better than the traditional K-means.

# Supporting Information

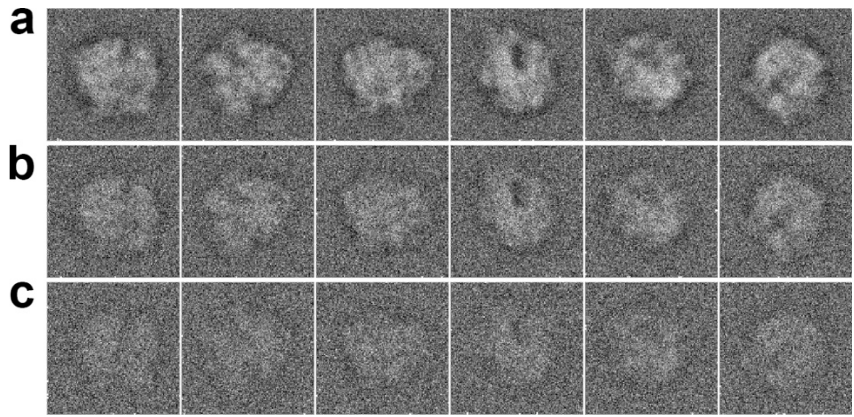

**S1 Fig**. **Simulated projections of *Escherichia coli* 70S ribosome.**

For each noise level, six orientations are shown. (a) NSR=1/3. (b) NSR = 1/10. (c) NSR=1/30.

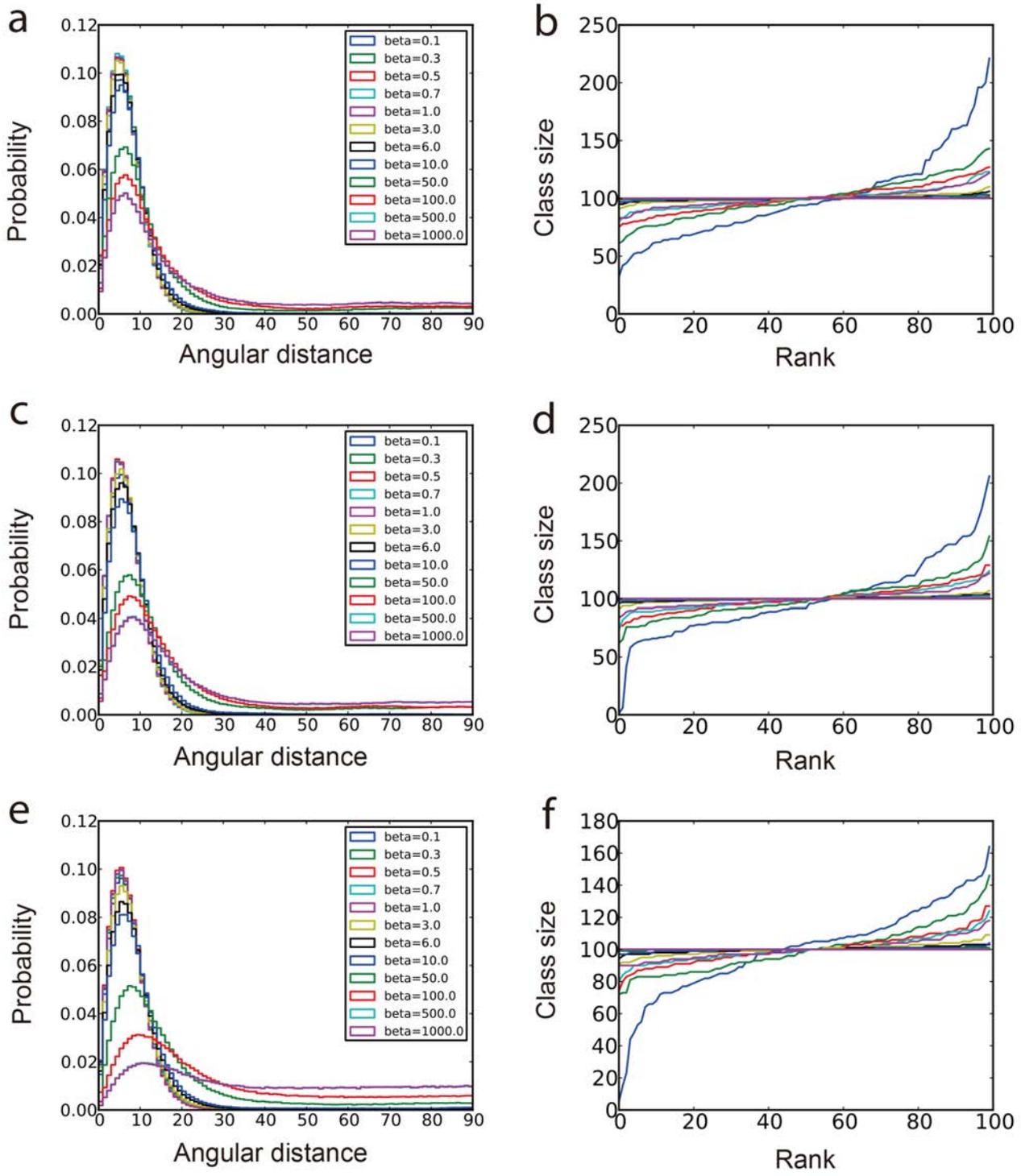

**S2 Fig. Comparison of classification results of ACK-means on simulated data with different β and at different noise levels.**

The First column (panels a, c and e) is the normalized histogram of angular distances. More accurate classification produces curve with higher peak concentrated at lower angular distance. The second column (panels b, d and f) is size of classes which is arranged in ascend order. The most balanced classification has a horizontal line in this plot. The experiments are conducted by MRA approach in SPARX. (a) and (b) NSR=1/3. (c) and (d) NSR = 1/10. (e) and (f) NSR=1/30.

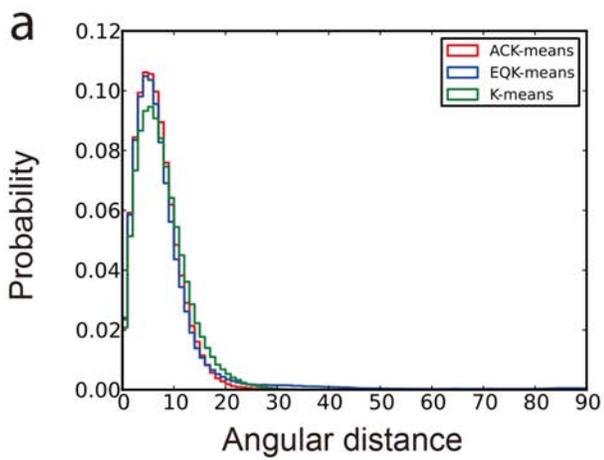
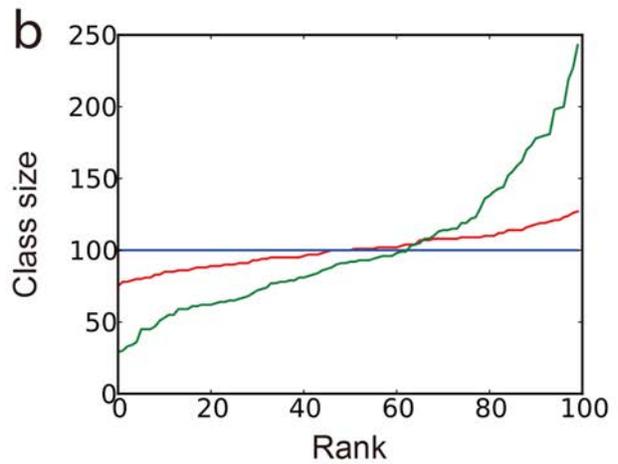
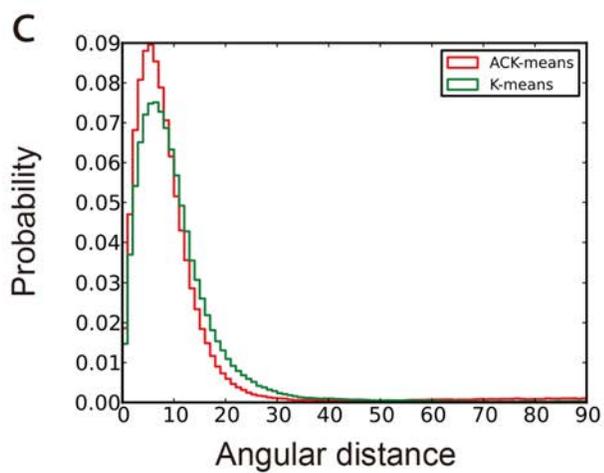
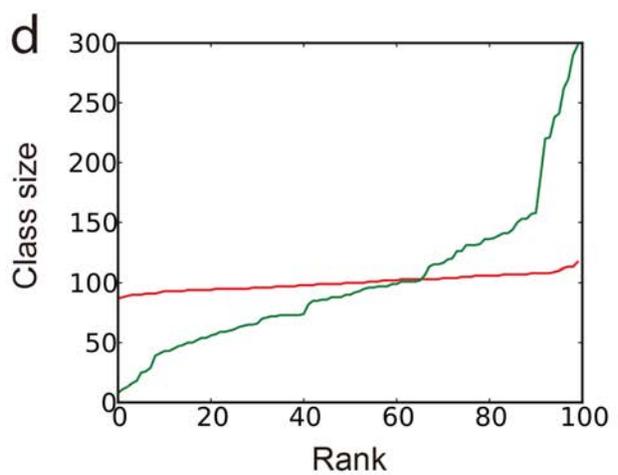
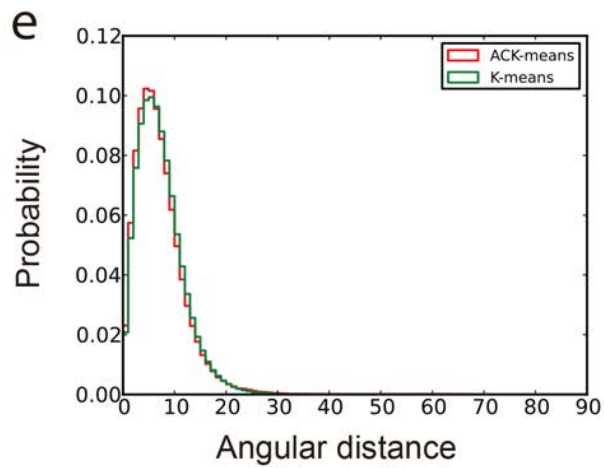
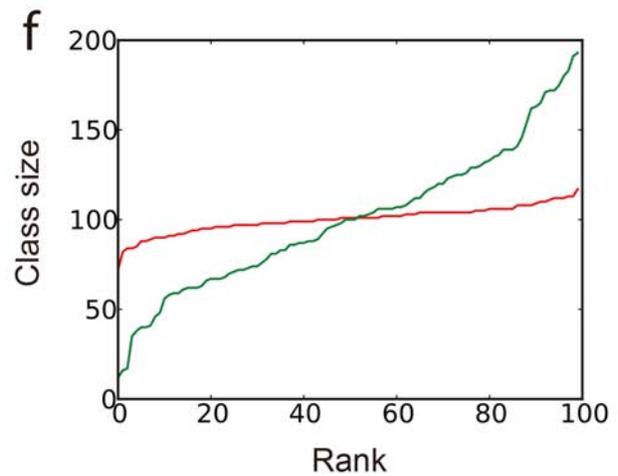

**S3 Fig**. **Comparison of classification results of simulated data with SNR = 1/3.**

The First column (panels a, c and e) is the normalized histogram of angular distances. More accurate classification produces curve with higher peak concentrated at lower angular distance. The second column (panels b, d and f) is size of classes which is arranged in ascend order. The most balanced classification has a horizontal line in this plot. (a) and (b) are for experiments using different clustering algorithms in MRA approach under SPARX. (c) and (d) are for experiments using different clustering algorithms in MRA/MSA approach under EMAN2. (e) and (f) are for experiments using different clustering algorithms in RFA approach under SPIDER. In all graphs, red curves present the results from ACK-means algorithm.

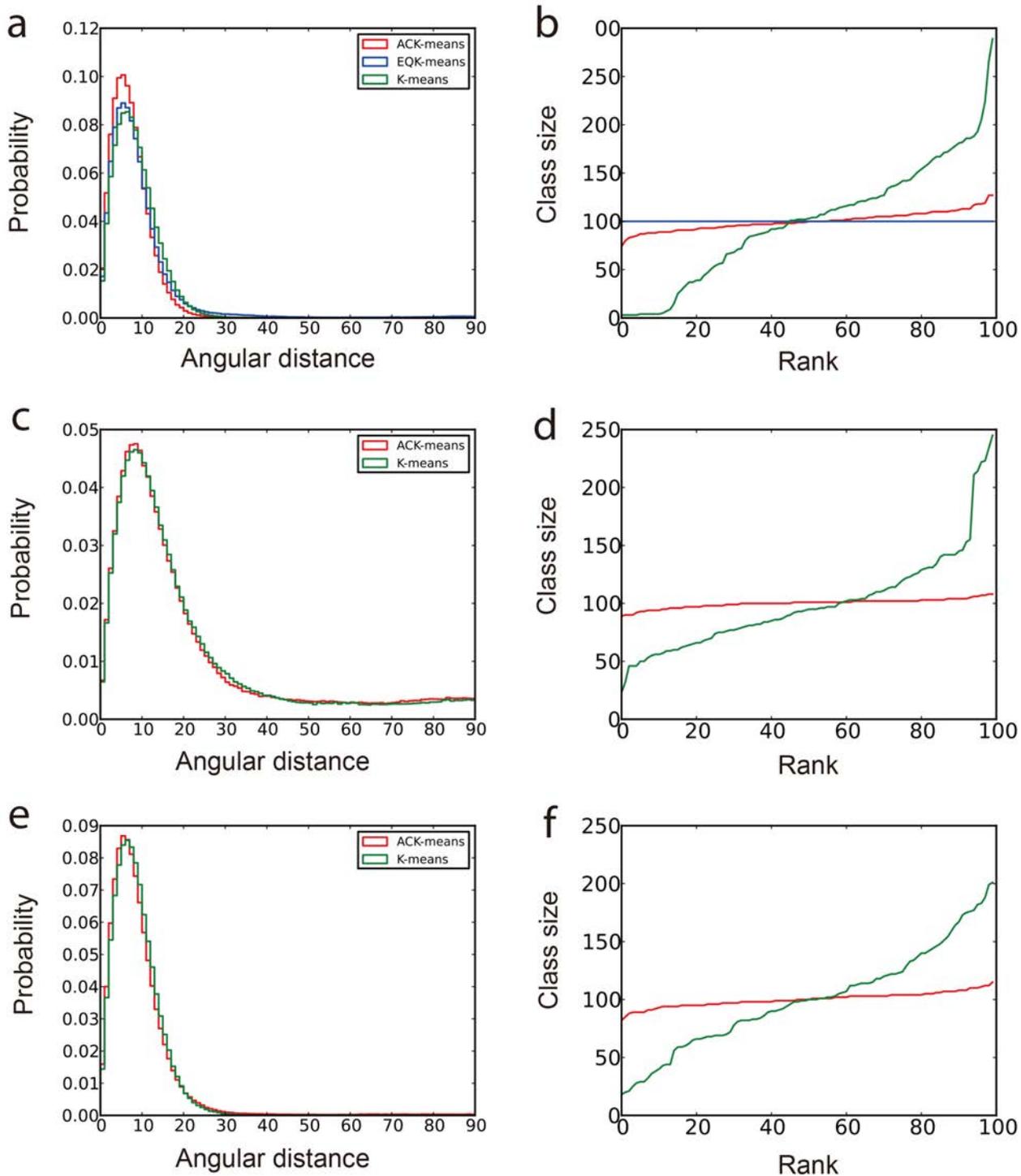

**S4 Fig**. **Comparison of classification results of simulated data with SNR = 1/30.**

The First column (panels a, c and e) is the normalized histogram of angular distances. More accurate classification produces curve with higher peak concentrated at lower angular distance. The second column (panels b, d and f) is size of classes which is arranged in ascend order. The most balanced classification has a horizontal line in this plot. (a) and (b) are for experiments using different clustering algorithms in MRA approach under SPARX. (c) and (d) are for experiments using different clustering algorithms in MRA/MSA approach under EMAN2. (e) and (f) are for experiments using different clustering algorithms in RFA approach under SPIDER. In all graphs, red curves present the results from ACK-means algorithm.